\documentclass{article}
\pdfoutput=1
\usepackage{graphicx}  
\usepackage{amsmath}   
\usepackage[compress]{cite}
\usepackage{amssymb}   
\usepackage{bm} 
\usepackage{dcolumn}
\usepackage{color}
\usepackage{mathrsfs}
\usepackage{amsfonts}
\usepackage{varioref}
\usepackage[caption=false]{subfig}
\RequirePackage[colorlinks,citecolor=blue,urlcolor=magenta,linkcolor=blue]{hyperref}
\input epsf

\labelformat{section}{Section #1} 
\labelformat{subsection}{Section #1} 
\labelformat{subsubsection}{Section #1}
\labelformat{subsubsubsection}{Section #1}
\labelformat{equation}{Eq.~(#1)} 
\labelformat{figure}{Fig.~#1} 
\labelformat{subfigure}{Fig.~\thefigure#1} 
\labelformat{table}{Tab.~#1} 
\labelformat{appendix}{Appendix #1}

\title{A critical analysis of modulus stabilization in a higher dimensional F(R) gravity}

\author{Indrani Banerjee$^{1}$\footnote{banerjeein@nitrkl.ac.in}~,
Tanmoy Paul$^{2,3}$\footnote{pul.tnmy9@gmail.com}~
and Soumitra SenGupta$^{4}$\footnote{tpssg@iacs.res.in} \\
\small{$^{1}$Department of Physics and Astronomy, National Institute of Technology, Rourkela-769008, India }\\
\small{$^{2}$ Department of Physics, Chandernagore College, Hooghly - 712 136, India}\\
\small{$^{3}$ International Laboratory for Theoretical Cosmology, TUSUR, 634050 Tomsk, Russia}\\
\small{$^{4}$ School of Physical Sciences, Indian Association for the Cultivation of Science, Kolkata-700032, India}}

\date{}
\begin{document}

\maketitle

\begin{abstract}
An exact solution for the bulk 5-dimensional geometry is derived for F(R) gravity with 
non-flat de-Sitter 3-branes located at the $M_4 \times Z_2$ orbifold boundaries. The corresponding form of F(R) that leads to such an exact solution of the bulk metric is derived which turns out to have all positive integer powers of R.
In such a scenario the stability issue of the modulus (radion field) is analyzed critically for different curvature epochs in both Einstein and Jordan frames. The radion in the effective 4-d theory exhibits a phantom epoch making this model viable for a non-singular bounce. Simultaneous resolution of the gauge hierarchy problem is exhibited through the resulting stable vaue of the radion field in the effective $3+1$ dimensional theory.

\end{abstract}


\section{Introduction}
The huge disparity between the electroweak scale and the Planck scale has given birth to the gauge hierarchy problem resulting in large radiative corrections to the Higgs mass. Extra dimensions, originally invoked to unify the known forces, have played an instrumental role in providing a resolution to the same \cite{Antoniadis:1990ew, Horava:1995qa, Horava:1996ma, PhysRevD.54.R3693, Kakushadze:1998wp, ArkaniHamed:1998rs,  Antoniadis:1998ig, PhysRevD.59.086004, Randall:1999ee, Randall:1999vf, Lykken:1999nb, Kaloper:1999sm}. In particular, the warped braneworld model proposed by Randall-Sundrum (RS) has been a subject of interest for a long time since it resolves the fine-tuning problem without introducing any new arbitrary scale in the theory \cite{Randall:1999ee}. The RS model comprises of a 5-dimensional AdS bulk with the extra dimension $\varphi$ possessing an $S_1 /Z_2$ orbifold symmetry. Two 3-branes are embedded in the bulk spacetime at the orbifold fixed point $\varphi=0$ (hidden/Planck brane) and $\varphi=\pi$ (visible/TeV brane). The resolution of the fine-tuning problem hinges crucially on the interbrane separation which gives rise to the requisite warping of the Higgs mass from the Planck brane to the TeV brane. Stabilizing the distance between the branes that gives rise to the necessary warping therefore becomes a pressing issue.
It is important to note that the distance between the two branes may fluctuate in general which introduces a new scalar field in the theory, 
the so-called radion or the modulus. This is an important prediction of the RS scenario as the radion assumes TeV scale mass and couplings 
to the Standard Model fields making its search viable in the collider experiments \cite{GOLDBERGER2000275,Csaki:2000zn,Paul:2016itm,Mitra:2017run}.

Goldberger \& Wise \cite{Goldberger:1999uk,GOLDBERGER2000275} proposed a mechanism to stabilize the modulus 
by introducing a scalar field in the bulk and achieved the warping required for resolution of the hierarchy 
problem by suitably adjusting the boundary values of the bulk scalar on the two 3-branes. 
They however did not take into account the modification induced in the bulk metric ( i.e the back-reaction ) 
due to the presence of the 5-dimensional stabilizing scalar field. Moreover, the origin of the bulk scalar remained unexplored 
in the Goldberger \& Wise scenario. The first issue was addressed by Csaki et al. \cite{Csaki:2000zn} who worked out an exact solution 
of the back-reacted metric due to the presence of the bulk scalar by assuming a mass term and a quartic self coupling of the scalar field. 
One of the  aim of the present work is to provide a plausible explanation of the second issue which pertains to the origin of the bulk scalar field. 

In the aforesaid higher dimensional scenarios the bulk curvature is of the order of the Planck scale where it is natural 
to introduce the higher curvature corrections to Einstein gravity \cite{Nojiri:2010wj,Nojiri:2017ncd}. In this regard, we consider the bulk to be governed 
by $F(R)$ gravity which provides the simplest modification to the Einstein-Hilbert action to include higher curvature terms. 
It is well-known that a conformal transformation of the metric maps $F(R)$ gravity to the scalar-tensor theory where the bulk 
higher curvature degrees of freedom manifests itself through a scalar field in the Einstein action. In this work we explore the 
prospect of this scalar field in stabilizing the modulus.

Attempts to stabilize the radion using higher curvature corrections to Einstein gravity have been explored 
previously in \cite{Chakraborty:2016gpg,Das:2017htt,Elizalde:2018rmz,Chakraborty:2014zya,Anand:2014vqa} but there the 3-branes were considered to be \emph{flat} . In all the earlier braneworld scenarios  studied by Randall \& Sundrum \cite{Randall:1999ee}, Goldberger \& Wise \cite{Goldberger:1999uk,GOLDBERGER2000275} or Csaki et al \cite{Csaki:2000zn} , it was assumed that the two 3-branes at the two orbifold fixed points  are flat with  Einstein gravity in the bulk. A flat braneworld arises when the bulk induced cosmological constant on the brane exactly cancels the brane tension resulting into a flat 3-brane endowed with a Minkowski metric. \cite{Sasaki:1999mi}. It would therefore be natural to explore a more realistic scenario  where the 3-branes are  non-flat branes which may arise from some imbalance between the two thereby rendering the branes to assume a de Sitter (dS) or anti-de Sitter (AdS) character. A de Sitter brane is particularly interesting since it can explain the observed accelerated expansion of the universe \cite{SupernovaCosmologyProject:1998vns,1538-3881-116-3-1009}.

Warped braneworld model with non-flat branes and bulk Einstein gravity was first explored in \cite{Das:2007qn} 
where it was shown that introducing non-flatness leads to a modified warping between the two branes which in turn can 
successfully address the gauge hierarchy issue (irrespective  of their dS or AdS nature). 
In the event of a vanishing brane cosmological constant the RS scenario is retrieved . 
Henceforth we will focus on dS branes since these are more relevant from the observational 
perspective \cite{SupernovaCosmologyProject:1998vns,1538-3881-116-3-1009}. When the interbrane separation is allowed to 
vary in the non-flat warped braneworld scenario (with bulk Einstein gravity) it can be shown that a radion potential is 
naturally generated at the level of the 4-dimensional effective action which in turn can successfully stabilize the modulus 
\cite{Banerjee:2017jyk}. The potential has an inflection point and by suitably adjusting the value of the brane cosmological 
constant the gauge hierarchy problem can also be resolved at the stable point. It was further shown that the radion kinetic term in 
such a scenario has a non-canonical coupling which  exhibits a phantom character for certain values of the modulus. 
This makes it a suitable candidate to explain several early universe scenarios \cite{Banerjee:2018kcz,Banerjee:2020uil,Das:2017jrl} 
(for bouncing cosmology and the violation of energy condition, see \cite{Brandenberger:2016vhg,Odintsov:2020zct}). 
Particularly, in a FRW background the negative kinetic term of the radion leads to a violation of the null energy condition making 
it viable for a bouncing universe \cite{Banerjee:2020uil}. The radion potential vanishes as the brane cosmological constant goes 
to zero (i.e. flat branes) and the kinetic term becomes canonical, in agreement with previous studies \cite{GOLDBERGER2000275}.

The goal of the present work is to investigate the non-flat warped braneworld scenario in purview of bulk $F(R)$ gravity. With bulk curvature as high as the Planck scale such higher curvature corrections to the bulk gravity action is a natural choice . By performing a conformal transformation of the metric we move to the Einstein frame where a bulk scalar naturally emerges from the higher curvature degrees of freedom with a potential whose properties depend on the form of $F(R)$. 
We allow the non-flat branes to be dynamical and derive the 4-dimensional radion effective action. Once again, the radion inherits a potential and a non-canonical kinetic coupling which can be attributed both to bulk $F(R)$ and the non-flat character of the branes. The role of such a potential in addressing the modulus stabilization and the gauge-hierarchy issue forms the main subject of this work.

The paper is organized as follows: In \ref{sectionFR} we discuss the 5-D bulk $F(R)$ action and its transformation to the Einstein frame. We solve the gravitational and scalar field equations in \ref{S2-1} and derive an exact form of the warp factor taking into account the backreaction due to the bulk scalar. With this warp factor the forms of the scalar potential and bulk $F(R)$ gravity are derived in \ref{S2-2} and \ref{S2-3} respectively. \ref{S3} is dedicated in deriving the radion effective action where the 4-dimensional Ricci scalar inherits a non-minimal coupling with the modulus. Such a coupling is removed in \ref{S3-2} by a second conformal transformation to the Einstein frame. Subsequently, the role of the radion effective potential in the context of modulus stabilization is addressed. We conclude with a summary of our results with some scope for future work in \ref{Conclusion}.

\section{Five dimensional warped braneworld scenario with bulk $F(\mathbf{R})$ gravity}
\label{sectionFR}
In the present work, we consider a five dimensional warped spacetime with two 3-brane scenario, with $S^1/Z_2$ orbifold symmetry along the extra spatial 
dimension $\varphi$ \cite{Randall:1999ee}. The two 3-branes are located at $\varphi = 0$ (hidden brane) and at 
$\varphi = \pi$ (visible brane) respectively, such that the latter one is identified with our four dimensional visible universe. In such warped braneworld scenario, since the bulk curvature is of the order of Planck scale, one may expect the bulk action to be governed by higher curvature gravity, e.g. $F(\mathbf{R})$ gravity which we consider in this work.
 
\begin{eqnarray}
 S_{bulk} = \int d^5x \sqrt{-\mathbf{G}}\bigg[\frac{F(\mathbf{R})}{2\kappa^2}\bigg] - \int d^4 x \sqrt{-\mathbf{g_{vis}}}~\lambda_{vis} ~\delta(\varphi-\pi) d\varphi- \int d^4 x \sqrt{-\mathbf{g_{hid}}} ~\lambda_{hid}~ \delta(\varphi) d\varphi
 \label{transformation1}
\end{eqnarray}
where $\mathbf{G}$ is the determinant of the five dimensional metric $\mathbf{G_{AB}}$ ($A$, $B$ runs from $0$ to $4$), $\mathbf{R}$ is the Ricci scalar constructed from $\mathbf{G_{AB}}$ and 
$\frac{1}{4\kappa^2} = M^{3}$ with $M$ the five dimensional Planck mass. The branes are endowed with constant energy densities, viz $\lambda_i$ ($i=vis(hid)$ for the visible (hidden) brane respectively.) 

By introducing an auxiliary field $\mathcal{A}(x)$, the action (\ref{transformation1}) can be rewritten in the so-called Jordan frame,
\begin{eqnarray}
\hspace{-0.6cm}
 S_{bulk} = \int d^5x \sqrt{-\mathbf{G}}\frac{1}{2\kappa^2}\bigg[F'(\mathcal{A})(\mathbf{\mathbf{R}}-\mathcal{A}) + F(\mathcal{A})\bigg] - \int d^4 x \sqrt{-\mathbf{g_{vis}}}~\lambda_{vis} ~\delta(\varphi-\pi)d\varphi - \int d^4 x \sqrt{-\mathbf{g_{hid}}} ~\lambda_{hid}~ \delta(\varphi)d\varphi 
 \label{transformation2}
\end{eqnarray}
The variation of this action with respect to the auxiliary field $\mathcal{A}(x)$ leads to $\mathcal{A} = \mathbf{R}$, which finally results in the original action (\ref{transformation1}). Moreover, the action (\ref{transformation2}) 
can be mapped into the Einstein frame by applying the following conformal transformation to the metric $\it{\mathbf{G_{AB}}}(x)$ in \ref{transformation1} \cite{PhysRevD.65.023521,RevModPhys.82.451,DeFelice2010,PhysRevD.92.026008,0264-9381-14-12-010,BARROW1988515,1475-7516-2005-04-014,PhysRevD.76.084039,1475-7516-2013-10-040},
\begin{equation}
 \mathbf{G_{AB}}(x) = e^{-[\sqrt{\frac{1}{3}}\kappa(\Psi-\Psi_P)]} \tilde{G}_{AB}(x)
 \label{transformation3}
\end{equation}
where $\Psi(x)$ is the conformal factor which is related to the auxiliary field as $F'(\mathcal{A}) = e^{\sqrt{\frac{3}{4}}\kappa(\Psi-\Psi_P)}$ and $\Psi_P$ is a constant whose significance will be clear in the next few sections. Due to the above conformal 
transfromation along with the mentioned relation among $\Psi$ and $F'(\mathcal{A})$, the bulk action in \ref{transformation1} can be mapped to the following scalar-tensor action in the Einstein frame,
\begin{align}
\hspace{-2.4cm}
^{(5)}S&=\int d^5x \sqrt{-\tilde{G}}\bigg[\frac{\tilde{R}}{2\kappa^2} - \frac{1}{2}\tilde{G}^{AB}\partial_{A}\Psi \partial_{B}\Psi - \tilde{ V}_E(\Psi)\bigg] + \int d^4 x \sqrt{-\tilde{g}_{vis}} \frac{\kappa}{2\sqrt{3}} \partial_\varphi \Psi\delta(\varphi-\pi) d\varphi +  \int d^4 x \sqrt{-\tilde{g}_{hid}} \frac{\kappa}{2\sqrt{3}} \partial_\varphi \Psi\delta(\varphi) d\varphi
\nonumber \\
&- \int d^4 x \sqrt{-\tilde{g}_{vis}}~\lambda_{vis} ~\rm exp\bigg\lbrace-{{\frac{2}{\sqrt{3}}}\kappa(\Psi-\Psi_P)}\bigg\rbrace ~\delta(\varphi-\pi) d\varphi- \int d^4 x \sqrt{-\tilde{g}_{hid}} ~\lambda_{hid}~\rm exp\bigg\lbrace -{{\frac{2}{\sqrt{3}}}\kappa(\Psi-\Psi_P)}\bigg\rbrace~ \delta(\varphi)d\varphi
 \label{transformation4}
\end{align}
where $\tilde{R}$ is constructed with the Einstein frame metric $\tilde{G}_{AB}$ and the potential due to the canonical scalar field $\Psi$ is given by,
\begin{eqnarray}
 \tilde{V}_E(\Psi) = \frac{1}{2\kappa^2}\left[\frac{\mathcal{A}}{F'(\mathcal{A})^{2/3}} - \frac{F(\mathcal{A})}{F'(\mathcal{A})^{5/3}}\right]~~.
 \label{bulk scalar potential}
\end{eqnarray}
Thus, the higher curvature 
degree of freedom manifests itself as a scalar field degree of freedom $\Psi$ with the potential $\tilde{V}_E(\Psi)$, which actually depends on the form of $F(\mathbf{R})$.
We further note that the conformal transformation of the Jordan frame metric $\mathbf{G_{AB}}$ gives rise to a surface term (which manifests as the second and the third term in \ref{transformation4}), while the brane tensions $\lambda_{vis}$ and $\lambda_{hid}$ (third and fourth term of \ref{transformation4}) also get modified suitably due to the conformal transformation.

We noted that $F(\mathbf{R})$ gravity can be equivalently mapped to scalar-tensor gravity with a suitable scalar potential where $\Psi$ denotes the bulk scalar field which originates from the higher curvature degrees of freedom, with the potential $\tilde{V}_E(\Psi)$ which in turn depends on the form of $F(\mathbf{R})$. The action of the model in the scalar-tensor frame is given by \ref{transformation4} which can also be written as,
\begin{align}
&^{(5)} S = \int d^4x \int_{-\pi}^{\pi}d\varphi \sqrt{-\tilde{G}}\left[2M^3\tilde{R} - \Lambda - \frac{1}{2}\tilde{G}^{AB}\partial_{A}\Psi\partial_{B}\Psi - \tilde{V}(\Psi)\right] + \int d^4 x \sqrt{-\tilde{g}_{vis}} \frac{2}{\sqrt{3}\kappa} \partial_\varphi \Psi\delta(\varphi-\pi) d\varphi 
\nonumber \\
&+  \int d^4 x \sqrt{-\tilde{g}_{hid}} \frac{2}{\sqrt{3}\kappa} \partial_\varphi \Psi\delta(\varphi) d\varphi - \int d^4 x \sqrt{-\tilde{g}_{vis}}~\lambda_{vis} ~\rm exp\bigg\lbrace-{{\frac{2}{\sqrt{3}}}\kappa(\Psi-\Psi_P)}\bigg\rbrace~\delta(\varphi-\pi) d\varphi  \nonumber \\
&- \int d^4 x \sqrt{-\tilde{g}_{hid}} ~\lambda_{hid}~\rm exp\bigg\lbrace-{{\frac{2}{\sqrt{3}}}\kappa(\Psi-\Psi_P)}\bigg\rbrace~ \delta(\varphi)d\varphi
\label{5D action}
\end{align}
where $\Lambda=-24M^3k^2$ represents the minima of the potential $\tilde{V}_E(\Psi)$ such that $\tilde{V}_E(\Psi)=\tilde{V}(\Psi) + \Lambda$, and can be interpreted as the bulk cosmological constant. \\

Varying the action with respect to $\tilde{G}^{AB}$ and $\Psi$ lead to the gravitational and the scalar field 
equations as follows:
\begin{align}
 \sqrt{-\tilde{G}}\left[\tilde{R}_{AB} - \frac{1}{2}\tilde{R}~\tilde{G}_{AB}\right]&=-\frac{1}{4M^3}\sqrt{-\tilde{G}}\left[\Lambda~ \tilde{G}_{AB} - \mathcal{L}_\mathrm{\Psi}\tilde{G}_{AB} 
 - \partial_{A}\Psi\partial_{B}\Psi\right] \nonumber \\
& - \sqrt{-\tilde{g}_\mathrm{hid}}\frac{V_{hid}}{4M^3}~~\tilde{g}_{\mu\nu}^{(hid)}\delta^{\mu}_{A}\delta^{\nu}_{B}
 \delta(\varphi) - \sqrt{-\tilde{g}_\mathrm{vis}}\frac{V_{vis}}{4M^3}~~\tilde{g}_{\mu\nu}^{(vis)}\delta^{\mu}_{A}\delta^{\nu}_{B}
 \delta(\varphi - \pi)
 \label{gravitational equation}
\end{align}
and
\begin{eqnarray}
 \Box{\Psi} = \frac{1}{\sqrt{-\tilde{G}}}\left(\sqrt{-\tilde{G}}~\partial^{A}\Psi\right)_{,A}
 \label{scalar field equation}
\end{eqnarray}
respectively, where $\mathcal{L}_\Psi$ is the Lagrangian due to the scalar field, $\Box$ denotes the d'Alembertian operator constructed from $\tilde{G}_{AB}$ while the effective brane tensions $V_{vis}$ and $V_{hid}$ are respectively given by,
\begin{align}
V_{hid} = \lambda_\mathrm{hid}e^{-\frac{2}{\sqrt{3}}\kappa\lbrace \Psi(0)-\Psi_P\rbrace} -\frac{2}{\sqrt{3}}\kappa ~ \partial_\varphi\Psi|_{\varphi=0}
\label{Vhid}
\end{align}
\begin{align}
V_{vis} = \lambda_\mathrm{vis}e^{-\frac{2}{\sqrt{3}}\kappa \lbrace\Psi(\pi)-\Psi_P\rbrace} -\frac{2}{\sqrt{3}}\kappa ~ \partial_\varphi\Psi|_{\varphi=\pi}
\label{Vvis}
 \end{align}

\subsection{Solution of the field equations}
\label{S2-1}
In this section we derive the solution of the field equations discussed earlier.
At this stage, we assume that the following metric ansatz solves the equations \ref{gravitational equation} and \ref{scalar field equation},
\begin{eqnarray}
 ds^2 = e^{-2A(\varphi)}\tilde{g}_{\mu\nu}(x)dx^{\mu}dx^{\nu} + r_c^2d\varphi^2~~,
 \label{1}
\end{eqnarray}
where $A(\varphi)$ is known as the warp factor that depends only on the extra dimensional coordinate $\varphi$ 
and obeys the normalization condition $A(\varphi = 0) = 1$. Also note that the Roman capitalized indices denote the bulk coordinates while the brane coordinates are denoted by the Greek indices.  
Moreover in \ref{1} $r_c$ represents the separation between the branes along the constant $x^{\mu}$ (strictly, the 
brane separation along the constant $x^{\mu}$ is $ \pi r_c$). The induced metric on the branes come as: 
$\tilde{g}_{\mu\nu}^{(hid)} = \tilde{g}_{\mu\nu}$ and $\tilde{g}_{\mu\nu}^{(vis)} = e^{-2A(\pi)}\tilde{g}_{\mu\nu}$ respectively. 
Using the above ansatz \ref{1}, the Ricci tensor and the Ricci scalar turn out to be,
\begin{eqnarray}
 \tilde{R}_{\mu\nu}&=&\tilde{R}_{\mu\nu}^{(4)} + \frac{1}{r_c^2}e^{-2A}A''(\varphi)\tilde{g}_{\mu\nu} - \frac{4}{r_c^2}e^{-2A}A'(\varphi)^2~\tilde{g}_{\mu\nu}\label{1-a}\\
 \tilde{R}&=&\tilde{R}^{(4)}e^{2A} + \frac{8}{r_c^2}A''(\varphi) - \frac{20}{r_c^2}A'(\varphi)^2~~,
 \label{1-b}
\end{eqnarray}
where $R_{\mu\nu}^{(4)}$ and $R^{(4)}$ are the four dimensional Ricci tensor and Ricci scalar respectively constructed from $\tilde{g}_{\mu\nu}(x)$. 
Consequently the ($\mu,\nu$) and ($\varphi,\varphi$) component of Eq.(\ref{gravitational equation}) take the following forms:
\begin{eqnarray}
 \tilde{R}_{\mu\nu}^{(4)} - \frac{\tilde{R}^{(4)}}{2}\tilde{g}_{\mu\nu} = \left[\frac{3A''}{r_c^2} - \frac{6A'^2}{r_c^2} - \frac{\Lambda}{4M^3} 
 - \frac{\Psi'^2}{8M^3r_c^2} - \frac{\tilde{V}(\Psi)}{4M^3} - \frac{V_{vis}}{4M^3} \delta(\varphi-\pi) - \frac{V_{hid}}{4M^3} \delta(\varphi) \right]e^{-2A(\varphi)}\tilde{g}_{\mu\nu}
 \label{gr eqn 1}
\end{eqnarray}
and
\begin{eqnarray}
 -\frac{1}{2}\tilde{R}^{(4)}e^{2A}r_c^2 + 6A'(\varphi)^2 = - \frac{\Lambda}{4M^3}r_c^2 + \frac{\Psi'^2}{8M^3} - \frac{\tilde{V}(\Psi)}{4M^3}r_c^2~~.
 \label{gr eqn 2}
\end{eqnarray}
We assume the bulk scalar $\Psi$ to depend only on $\varphi$ while $\tilde{g}_{\mu\nu}$ depends only on the brane coordinates. Dividing \ref{gr eqn 1} by $\tilde{g}_{\mu\nu}$ on both sides we note that the left hand side depends only on the four dimensional coordinate $x^{\mu}$ while the right hand 
side has the only dependence on $\varphi$. Therefore both the sides of \ref{gr eqn 1} must be individually equal to a constant, say $-\Omega$, 
\begin{eqnarray}
 R_{\mu\nu}^{(4)} - \frac{1}{2}R^{(4)}g_{\mu\nu}&=&-\Omega g_{\mu\nu}~~,\label{gr eqn 3}\\
 \left[\frac{3A''}{r_c^2} - \frac{6A'^2}{r_c^2} - \frac{\Lambda}{4M^3} 
 - \frac{\Psi'^2}{8M^3r_c^2} - \frac{\tilde{V}(\Psi)}{4M^3} - \frac{V_{vis}}{4M^3} \delta(\varphi-\pi) - \frac{V_{hid}}{4M^3} \delta(\varphi) \right]e^{-2A(\varphi)}&=&-\Omega~~,
 \label{gr eqn 4}
\end{eqnarray}
where $\Omega$ is the constant of separation. It may be observed from \ref{gr eqn 3} that the $\Omega$ can be treated as 
the induced cosmological constant on the brane, which in turn indicates that the branes in the present context are $curved$ mainly due to the presence 
of the non-zero cosmological constant. The trace of \ref{gr eqn 3} immediately leads to $R^{(4)} = 4\Omega$ which when substituted in \ref{gr eqn 2} leads to the following gravitational field
equations,
\begin{eqnarray}
 3A''(\varphi)&=&\Omega r_c^2e^{2A(\varphi)} + \frac{\Psi'^2}{4M^3} + \frac{V_{vis}}{4M^3} \delta(\varphi-\pi) + \frac{V_{hid}}{4M^3} \delta(\varphi)~~,\label{2}\\
  6A'(\varphi)^2&=&-\frac{\Lambda}{4M^3}r_c^2 - \frac{\tilde{V}(\Psi)}{4M^3}r_c^2 + \frac{\Psi'^2}{8M^3}  + 
  2 \Omega r_c^2e^{2A(\varphi)}~~,
 \label{3}
\end{eqnarray}
Furthermore due to the warped metric (\ref{1}), the scalar field equation of motion \ref{scalar field equation} can be expressed as,
\begin{eqnarray}
 \frac{1}{r_c^2}\Psi''(\varphi) = \frac{4}{r_c^2}A'(\varphi)\Psi'(\varphi) + \frac{d\tilde{V}}{d\Psi} + \frac{dV_{vis}}{d\Psi} \delta(\varphi-\pi) + \frac{dV_{hid}}{d\Psi} \delta(\varphi) ~~.
 \label{scalar field eqn 1}
\end{eqnarray}
where the prime denotes $\frac{d}{d\varphi}$, i.e the differentiation with respect to the extra dimensional angular coordinate. 
As a whole, \ref{2}, \ref{3} and \ref{scalar field eqn 1} are the key equations of motion for the present warped braneworld scenario 
described by the action (\ref{5D action}). However it may be mentioned that out of the three equations two are independent, i.e.,
the scalar field equation can be derived from \ref{2} and \ref{3} and thus these two equations will be regarded as independent ones. 

\ref{2} and \ref{3} clearly indicate that both the bulk scalar field and the brane cosmological constant 
contribute to the bulk spacetime curvature and hence the warp factor depends on the profile of the scalar field potential $V(\Psi)$. However, in the absence of the scalar field but non-flat branes, it can be shown that the following warp factor exactly solve \ref{2} and \ref{3},
\begin{eqnarray}
 e^{-A(\varphi)} = \omega\sinh\left[\ln{\frac{c_2}{\omega}} - kr_c\varphi\right]~~,
 \label{3-1}
\end{eqnarray}
when the brane cosmological constant $\Omega>0$ is considered, as observed in our present universe. In \ref{3-1} $\omega^2=\frac{\Omega}{3k^2}$ while $c_2 = 1 + \sqrt{1 + \omega^2}$ and $k = \sqrt{-\Lambda/\left(24M^3\right)}$. Further, when the interbrane separation is treated as a field (the so-called radion or the modulus), i.e., $r_c\equiv T(x)$, it can be shown that such non-flat branes give rise to a potential and a non-canonical kinetic term for the radion field at the level of 4-d effective action \cite{Banerjee:2017jyk,Banerjee:2018kcz,Banerjee:2020uil}. We will discuss this aspect in greater detail in the present context.

Another avenue to obtain an exact solution for the warp factor was investigated by Csaki et al.\cite{Csaki:2000zn} who showed that in the presence of bulk scalar with $Minkowskian$ brane (i.e for $\Omega = 0$), the above equations (\ref{2}, \ref{3} and \ref{scalar field eqn 1}) give an exact solution for the warp factor if the scalar potential $V(\Psi)$ is considered to have a specific combination of quadratic and quartic power of $\Psi$ where the corresponding coefficients are chosen accordingly.

However since we have both non-flat branes and bulk scalar field here, the potential 
considered in \cite{Csaki:2000zn} will not lead to a closed form of the warp factor $A(\varphi)$. Therefore in order to solve \ref{2} and \ref{3} where both the backreaction of 
the bulk scalar field and the brane cosmolgical constant are taken account, we consider the following ansatz for the bulk scalar $\Psi$,
\begin{eqnarray}
\Psi'(\varphi)^2 = 12M^3\gamma A''(\varphi)~~,
\label{4}
\end{eqnarray}
with $\gamma$ being a dimensionless parameter. Here it deserves mentioning that one can propose another ansatz which also leads to analytical solutions of $A(\varphi)$ and $\Psi(\varphi)$ like, $\Psi'(\varphi)^2 \propto e^{2A(\varphi)}$. However with such an ansatz, presence of the scalar field only modifies the brane cosmological constant $\omega$ keeping the form of the radion potential and the kinetic term unchanged in the 4-d effective action from what was obtained in \cite{Banerjee:2017jyk,Banerjee:2018kcz,Banerjee:2020uil}. Since such an ansatz does not yield any new feature in the radion effective action (compared to \cite{Banerjee:2017jyk,Banerjee:2018kcz,Banerjee:2020uil}) we will concentrate on \ref{4} in this work.

The parameter $\gamma$ in \ref{4} indicates the strength of the backreaction of $\Psi(\varphi)$ in the bulk geometry. Therefore, we will eventually see that for $\gamma = 0$, $\Psi(\varphi)$ acquires a constant value throughout the bulk such that it acts  like a bulk cosmological constant in the five dimensional spacetime. With the aforesaid ansatz, \ref{2} assumes the following form in the bulk,
\begin{eqnarray}
 A''(\varphi)(1-\gamma) = \frac{\Omega r_c^2}{3}e^{2A(\varphi)}=\omega^2~k^2r_c^2e^{2A(\varphi)}~~.
 \label{4a}
\end{eqnarray}
The above equation can be written as,
\begin{eqnarray}
 A''(\varphi) = \omega_m^2~k^2r_c^2e^{2A(\varphi)}
 \label{5}
\end{eqnarray}
where $\omega_m^2 = \omega^2/(1-\gamma)$. \ref{5} can be analytically solved for $A(\varphi)$ and given by,
\begin{eqnarray}
 e^{-A(\varphi)} = \omega_m\sinh\left[\Big|\ln{\frac{c_2}{\omega_m}} - kr_c|\varphi|\Big|\right]~~,
 \label{6}
\end{eqnarray}
with $c_2 = 1 + \sqrt{1 + \omega_m^2}$ and $k = \sqrt{-\Lambda/\left(24M^3\right)}$. Therefore for AdS bulk, i.e for $\Lambda < 0$, 
the quantity $k$ becomes real and consequently the five dimensional bulk spacetime gets exponentially warped; otherwise $k$ becomes 
imaginary and the spacetime metric gets an oscillating nature. However in order to resolve the gauge hierarchy problem, the spacetime needs to be 
exponentially warped and thus we take $\Lambda < 0$, i.e the bulk geometry is considered to be of AdS nature.

\subsection{The forms of bulk scalar field, its potential and the effective brane tensions}
\label{S2-2}
From the above solution of $A(\varphi)$ and using
\begin{eqnarray}
 A'(\varphi)&=&kr_c\coth\left[\Big|\ln{\frac{c_2}{\omega_m}} - kr_c|\varphi|\Big|\right] \label{6a}\\
 A''(\varphi)&=&\left(kr_c\right)^2 \sinh^{-2}\left[\Big|\ln{\frac{c_2}{\omega_m}} - kr_c|\varphi|\Big|\right]
 \label{6b}
\end{eqnarray}
we determine the potential for the bulk scalar field (i.e $\tilde{V}(\Psi)$) 
from \ref{3}, which assumes the following form,
\begin{eqnarray}
\tilde{ V}(\Psi(\varphi)) = -18\gamma M^3k^2 \sinh^{-2}\left[\Big|\ln{\frac{c_2}{\omega_m}} - kr_c|\varphi|\Big|\right]~~.
 \label{7}
\end{eqnarray}
It is evident from \ref{6b} that $A''(\varphi)$ is positive, and thus \ref{4a} demonstrates that for $\gamma < 1$, 
the brane cosmological constant becomes positive leading to de-Sitter branes, while for $\gamma > 1$, the branes appear to be anti de-Sitter. 
Since the observed universe is known to have a positive cosmological constant, we will concentrate mainly on 
the warped braneworld scenario with de-Sitter branes, i.e the case with $\gamma < 1$. 
In order to determine $\tilde{V} = \tilde{V}(\Psi)$ from \ref{7}, we need the solution of the bulk scalar field 
$\Psi = \Psi(\varphi)$, and for this purpose, we use the initial ansatz i.e \ref{4}. Plugging the solution of $A(\varphi)$ into 
\ref{4} yields,
\begin{eqnarray}
 \Psi'(\varphi)^2 = 12\gamma M^3k^2r_c^2~\sinh^{-2}\left[\Big|\ln{\frac{c_2}{\omega_m}} - kr_c|\varphi|\Big|\right]~~,
 \label{8}
\end{eqnarray}
which gives two possible roots of $\Psi'(\varphi)$. However we will take the negative root as $\Psi(\varphi)/M^{\frac{3}{2}}$ is considered to acquire 
larger values on the Planck brane compared to that in the TeV brane, in order to solve the gauge hierarchy problem. As a result,
\begin{eqnarray}
 \Psi'(\varphi) = -kr_c\sqrt{12\gamma M^3}~\sinh^{-1}\left[\Big|\ln{\frac{c_2}{\omega_m}} - kr_c|\varphi|\Big|\right]~~.
 \label{9}
\end{eqnarray}
Solving the above equation for $\Psi = \Psi(\varphi)$, we get
\begin{align}
 \Psi(\varphi) = \Psi_\mathrm{P} + \sqrt{12\gamma M^3}~
 \ln{\left\{\left(\frac{c_2 + \omega_m}{c_2 - \omega_m}\right)\tanh\left[\frac{|\Phi(\varphi)|}{2}\right]\right\}}~~,
 \label{10}
\end{align}
with $\Phi(\varphi) = \ln{\frac{c_2}{\omega_m}} - kr_c|\varphi|$ and $\Psi_\mathrm{P}$ denotes the bulk scalar field in the Planck brane, 
i.e $\Psi_\mathrm{P} = \Psi(\varphi = 0)$. 
Using \ref{10} it can be shown that,
\begin{align}
\tilde{V}(\Psi) = 18\gamma M^3k^2 \sinh^{2}\left[\frac{1}{\sqrt{12\gamma M^3}}\left(\Psi - \Psi_\mathrm{P}\right) 
 + \ln{\left(\frac{c_2 - \omega_m}{c_2 + \omega_m}\right)}\right]~~.
 \label{11}
\end{align}
As a whole, \ref{6} and \ref{10} denote the solutions of $A(\varphi)$ and $\Psi(\varphi)$ respectively, while the scalar field potential 
$\tilde{V}(\Psi)$ is reconstructed in \ref{11}. It is important to note that $\gamma<0$ is not allowed as this makes the bulk scalar field complex (\ref{10}), such that $\gamma$ lies in the range: $0<\gamma<1$.
Further from \ref{11} it may be observed that 
for $\gamma = 0$, $\tilde{V}(\Psi)$ vanishes while the bulk scalar field becomes constant throughout the bulk
(i.e., $\lim_{\gamma \to 0} \Psi(\varphi) = \Psi_\mathrm{P}$ from \ref{10}) and so the does the potential $\tilde{V}_E(\Psi)$.\\

In order to determine the effective brane tensions $V_{hid}$ and $V_{vis}$ we substitute the form of the ansatz \ref{4} in \ref{2} and then integrate  \ref{2} about $\varphi=0$ and $\varphi=\pi$ respectively. This yields,
\begin{align}
V_{hid}=24M^3k (1-\gamma)\Bigg[\frac{1+\frac{\omega_m^2}{c_2^2}}{1-\frac{\omega_m^2}{c_2^2}}\Bigg]
\label{VHid}
\end{align}
\begin{align}
V_{vis}=24M^3k (1-\gamma)\Bigg[\frac{\frac{\omega_m^2}{c_2^2}e^{kr_c\pi} + 1}{\frac{\omega_m^2}{c_2^2}e^{kr_c\pi} -1 }\Bigg]
\label{VVis}
\end{align}

\subsection{Form of bulk $F(\mathbf{R})$}
\label{S2-3}
Having reconstructed the scalar potential, now we would like to determine the form of $F(\mathbf{R})$ which gives rise to such a scalar potential. 
As mentioned earlier, the bulk scalar degrees of freedom can be encapsulated within the higher curvature $F(\mathbf{R})$ gravity. For the scalar 
potential $\tilde{V}_E(\Psi)$, the corresponding form of $F(\mathbf{R})$ in 5 dimension is given by \cite{Nojiri:2010wj},
\begin{eqnarray}
 F(R) = \frac{R}{2\kappa^2}\exp{\left[\frac{\kappa\left(\Psi(R) - \Psi_\mathrm{P}\right)}{\sqrt{3}}\right]} - \tilde{V}_E\left(\Psi(R)\right)
 \exp{\left[\frac{2\kappa\left(\Psi(R) - \Psi_\mathrm{P}\right)}{\sqrt{3}}\right]}~~,
 \label{form of F 1}
\end{eqnarray}
where $\Psi = \Psi(R)$ can be obtained by the following algebraic equation,
\begin{eqnarray}
 R = \exp{\left[\frac{\kappa\left(\Psi - \Psi_\mathrm{P}\right)}{\sqrt{3}}\right]}\left\{4\kappa^2 \tilde{V}_E(\Psi) + 2\kappa\sqrt{3}\tilde{V}'_E(\Psi)\right\}~~.
 \label{form of F 2}
\end{eqnarray}
Clearly, in order to obtain $\Psi = \Psi(R)$, one needs to invert \ref{form of F 2}. For this purpose, we use the 
condition $\kappa(\Psi - \Psi_\mathrm{P}) < 1$, i.e the scalar field $\Psi$ is considered to be sub-Planckian throughout the bulk. 
Such condition along with the $\tilde{V}(\Psi)$ in \ref{11} leads to the solution of $\Psi = \Psi(R)$ from \ref{form of F 2} as follows,
\begin{eqnarray}
 \frac{1}{\sqrt{3}}\kappa\left(\Psi - \Psi_\mathrm{P}\right) = \frac{R/\left(18\gamma k^2\right) - 2\Lambda/\left(9\gamma M^3k^2\right)}
 {\left(\frac{c_2 - \omega_m}{c_2 + \omega_m}\right)^2\left(1 + \frac{1}{2\sqrt{\gamma}}\right)\left(1 + \frac{1}{\sqrt{\gamma}}\right) 
 + \left(\frac{c_2 + \omega_m}{c_2 - \omega_m}\right)^2\left(1 - \frac{1}{2\sqrt{\gamma}}\right)\left(1 - \frac{1}{\sqrt{\gamma}}\right) 
 + \frac{2\Lambda}{9\gamma M^3k^2} - 2}~~,
 \label{form of F 3}
\end{eqnarray}
where we retain upto the leading order in $\kappa(\Psi - \Psi_\mathrm{P})$ and recall, $c_2 = 1 + \sqrt{1 + \omega_m^2}$. 
Plugging back the above solution of $\Psi(R)$ into \ref{form of F 1}, we obtain the final form of 
F(R) in the 5D bulk in the present context as,
\begin{eqnarray}
 F(R)&=&\frac{R}{2\kappa^2}
 \exp{\left\{\frac{1}{P\left(\gamma,\Lambda,\omega\right)}\left(\frac{R}{18\gamma k^2} - \frac{2\Lambda}{9\gamma M^3k^2}\right)\right\}}
 \left(1 - \frac{Q\left(\gamma,\Lambda,\omega\right)}{2P\left(\gamma,\Lambda,\omega\right)}\right)\nonumber\\ 
 &-&\Lambda\left(\frac{R/\left(18\gamma k^2\right) - 2\Lambda/\left(9\gamma M^3k^2\right)}{P\left(\gamma,\Lambda,\omega\right)}\right)^2~,
 \label{form of F 4}
\end{eqnarray}
where,
\begin{eqnarray}
 P\left(\gamma,\Lambda,\omega\right)&=&\left(\frac{c_2 - \omega_m}{c_2 + \omega_m}\right)^2\left(1 + \frac{1}{2\sqrt{\gamma}}\right)\left(1 + \frac{1}{\sqrt{\gamma}}\right) 
 + \left(\frac{c_2 + \omega_m}{c_2 - \omega_m}\right)^2\left(1 - \frac{1}{2\sqrt{\gamma}}\right)\left(1 - \frac{1}{\sqrt{\gamma}}\right) 
 + \frac{2\Lambda}{9\gamma M^3k^2} - 2\nonumber\\
 Q\left(\gamma,\Lambda,\omega\right)&=&\left(\frac{c_2 - \omega_m}{c_2 + \omega_m}\right)^2\left(1 + \frac{1}{\sqrt{\gamma}}\right) 
 + \left(\frac{c_2 + \omega_m}{c_2 - \omega_m}\right)^2\left(1 - \frac{1}{\sqrt{\gamma}}\right) - 2
 \label{P and Q}
\end{eqnarray}
respectively. 
The presence of the exponential term in the above expression reveals that the F(R) in the present context contains all positive integer powers of the Ricci 
scalar where the corresponding coefficients are connected by the parameter $\gamma$. In the present warped braneworld scenario where the bulk curvature is 
of the order of Planck scale, in particular $R \sim k^2$, such higher curvature terms are expected which play a crucial role in the spacetime geometry. Such a form of F(R) further corroborates the viability of the ansatz considered in \ref{4}. We note from \ref{10} that as $\gamma\to 0$, (i.e. the scalar field assumes the value $\Psi_P$ as in \ref{10}) $F(R)\to\frac{R}{2\kappa^2}-\Lambda$ (see \ref{form of F 1}).

\section{Deriving the 4-d effective action for the radion field}
\label{S3}
In this section we allow the interbrane distance to fluctuate slightly about its v.e.v $r_c$ such that it can be treated as a field, the so-called radion or the modulus, denoted by $T(x)$. The solution of the warp factor derived in \ref{S2-1} now becomes,
\begin{eqnarray}
 e^{-A(x^\mu,\varphi)} = \omega_m\sinh\left[\Big|\ln{\frac{c_2}{\omega_m}} - kT(x)|\varphi|\Big|\right]~~,
 \label{6-a}
\end{eqnarray}
 
In the last section we noted that the bulk action $^{(5)}{S}$ in the Einstein frame can be written as,
\begin{align}
^{(5)}{S}={S}_{g}+S_{\Psi} + {S}_{vis}+{S}_{hid} \label{Eq4}
\end{align}
where,
\begin{align}
{S}_{g}=\int_{-\infty} ^{\infty} d^4x \int_{-\pi} ^{\pi} d\varphi \sqrt{-\tilde{G}}(2M^3 {\tilde{ R}} - \Lambda ) \label{Eq5}
\end{align}

\begin{align}
S_{\Psi}=\int_{-\infty} ^{\infty} d^4x \int_{-\pi} ^{\pi} d\varphi \sqrt{-\tilde{G}}\bigg[-\frac{1}{2}\tilde{G}^{AB}\partial_{A}\Psi \partial_{B}\Psi - \tilde{ V}(\Psi)\bigg] 
\label{Eq5a}
\end{align}

\begin{align}
{S}_{vis}=\int_{-\infty}^{\infty} d^4x \int_{-\pi} ^{\pi} d\varphi \sqrt{-\tilde{g}_{vis}} (L_{vis} -V_{vis}) \delta(\varphi-\pi)  \label{Eq6}
\end{align}

\begin{align}
{S}_{hid}=\int_{-\infty}^{\infty} d^4x  \int_{-\pi} ^{\pi} d\varphi  \sqrt{-\tilde{g}_{hid}}(L_{hid} -V_{hid})\delta(\varphi)  \label{Eq7}
\end{align}
with $\tilde{R}$ the 5-dimensional Ricci scalar, $\sqrt{-\tilde{G}}=e^{-4A}T(x)\sqrt{-\tilde{g}}$ and $\Lambda=-24M^3k^2$ the bulk cosmological constant. For completeness we mention that the visible and hidden branes may contain matter Lagrangian given by $L_{vis}$ and $L_{hid}$ respectively which however can be ignored in the present context. Therefore, in what follows we will not consider $L_{vis}$ and $L_{hid}$.

We recall from the last section that the metric ansatz which solves the field equations assumes the form \ref{1} with $r_c$ replaced by $T(x)$ such that,

\begin{align}
 ds^2 = e^{-2A(x^\mu,\varphi)}\tilde{g}_{\mu\nu}(x)dx^{\mu}dx^{\nu} + T(x)^2d\varphi^2~~,
 \label{Eq8}
\end{align}

With the above ansatz the 5-d Ricci scalar $\tilde{R}$ can be written as,
\begin{align}
{\tilde{ R}}&= \tilde{R}{^{(4)}}e^{2A} + 8\frac{A^{\prime\prime}}{T(x)^2}-20\frac{{A^{\prime}}^2}{T(x)^2}-2e^{2A}\frac{(T(x)_,^{~\mu})_{;\mu}}{T(x)} +6e^{2A}(A_,^{~\nu})_{;\nu} \nonumber \\ &
-6e^{2A} A_,^{~\alpha} A_{,\alpha}+4e^{2A}\frac{T(x)_{,\rho}A^{~\rho}}{T(x)}  \label{Eq9}
\end{align}
with $\tilde{R}{^{(4)}}$ the 4-d Ricci scalar constructed from the metric $\tilde{g}_{\mu\nu}$ and ${A^{\prime}}$ represents derivative of $A$ with respect to the extra-coordinate $\varphi$.

The four terms of \ref{Eq9} contribute to the kinetic term of the radion field, in particular, we get,
\begin{align}
12M^3k\int_{-\pi}^{\pi} e^{-2A}|\varphi|d\varphi \int_\infty^\infty d^4 x\sqrt{-g}~ P(T(x),\varphi) 
\Big[kT(x)\varphi P(T(x),\varphi) - 1\Big]\partial_\mu T(x) \partial^\mu T(x) 
\label{Eq2}
\end{align}
where,
\begin{align}
P(T(x),\varphi)=\rm coth\Big(ln\frac{c_2}{\omega}-kT(x)|\varphi|\Big)
\label{Eq3}
\end{align} 
Eq. \ref{Eq2} above can also be written as,
\begin{align}
12M^3k\int_{-\pi}^{\pi} e^{-2A}|\varphi|d\varphi \int_\infty^\infty d^4 x\sqrt{-g}~ L(T(x)) \partial_\mu T(x) \partial^\mu T(x)
\label{Eq4}
\end{align}
where $L(T(x))$ is the coefficient of the kinetic term which at different regime determine the phantom and normal behavior of the 
radion field $T(x)$, in particular, $L(T(x)) > 0$ leads to the phantom regime of the radion field (recall, our metric signature is (-,+,+,+,+)). 
The form of $L(T(x))$ is,
\begin{align}
L(T(x))=P(T(x),\varphi) \Big[kT(x)|\varphi|P(T(x),\varphi) - 1\Big]
\end{align}
We note from \ref{Eq3} that $P(T(x),\varphi)>0$ when $\frac{\omega}{c_2}<e^{-kT(x)|\varphi|}$. 
In our case, $\omega/c_2<1$ and $0<\varphi<\pi$. In the event $\varphi=0$, $L(T(x)) = P(T(x),\varphi = 0)$ which is indeed positive for 
$\omega/c_2<1$ that is trivially satisfied. At $\varphi=\pi$, $P(T(x),\varphi)$ is positive when $\omega/c_2<e^{-kT(x)\pi}=\xi$. 
As a result, the radion field acquires a phantom regime in 
\begin{eqnarray}
 kT(x)\pi P(T(x),\varphi) > 1~~.
 \label{new1}
\end{eqnarray}
Due to the fact that $\omega/c_2 \ll 1$, the above condition can be approximated as $kT(x)\pi > 1$ or equivalently $\xi < e^{-1}$. 
Hence, the kinetic term of the radion field exhibits a phantom regime in 5-d when $\xi < e^{-1}$.

The brane tensions on the visible and hidden brane are obtained from \ref{VVis} and \ref{VHid} with $r_c$ replaced by $T(x)$, such that,
\begin{align}
V_{vis}=24M^3k(1-\gamma)\Bigg[\frac{\frac{\omega_m^2}{c_2^2}e^{2kT(x)\pi}+1}{\frac{\omega_m^2}{c_2^2}e^{2kT(x)\pi}-1}\Bigg] ~~~~~~~~~~~~~~~~~~~~{\rm and} ~~~~~~~~~~~~V_{hid}=24M^3k(1-\gamma)\Bigg[\frac{1+\frac{\omega_m^2}{c_2^2}}{1-\frac{\omega_m^2}{c_2^2}}\Bigg]  \label{Eq10}
\end{align}

Using \ref{Eq8}, \ref{Eq9} and \ref{Eq10} it can be shown that the effective action in 4-d is given by,
\begin{align}
S^J = \frac{2M^3}{k} \int_{-\infty} ^{\infty} d^4x \sqrt{-\tilde{g}}~h(\xi)~\tilde{R}{^{(4)}}  -\frac{6M^3c_2^2}{k}\int_{-\infty}^{\infty} d^{4}x \sqrt{-\tilde{g}}~ \frac{1}{2} \mathcal{G}(\xi) ~\partial_\mu \xi\partial^\mu\xi -2M^3 k \int_{-\infty} ^{\infty} d^4x \sqrt{-\tilde{g}} \mathcal{V}(\xi)
\label{Eq11}
\end{align}
where, $\xi= exp \lbrace-k\pi T(x) \rbrace$ is the dimensionless radion field. Henceforth, we will denote the radion field by $\xi$ which has the following physically allowed range: $0\leq \xi \leq 1$ as $T(x)$ cannot be negative since it represents the interbrane distance.

The non-minimal coupling of the radion to the Ricci scalar is obtained by integrating $\int_{-\pi}^\pi e^{-2A} d\varphi$, such that,
\begin{equation}
h(\xi)=\left\{\frac{c_2^2}{4}+\omega_m^2 \ln\left(\xi\right)+\frac{\omega_m^4}{4c_2^2}\left(\frac{1}{\xi^2}\right)-\frac{\omega_m^4}{4c_2^2}-\frac{c_2^2}{4}\xi^2 \right\}
\label{Eq12}
\end{equation}

The non-canonical coupling to the kinetic term $\mathcal{G}(\xi)$ is obtained by multiplying $e^{-4A}$ originating from $\sqrt{-\tilde{G}}$ to the last four terms of \ref{Eq9} and integrating with respect to $\varphi$ such that,
\begin{align}
\mathcal{G}(\xi)=  1-\frac{\omega_m^4}{c_2^4}\frac{1}{\xi^4}+\frac{4}{3}\frac{\omega_m^2}{c_2^2}\frac{1}{\xi^2}{ \rm ln} ~\xi 
\label{Eq13}
\end{align}

Finally, the potential of the radion field $\mathcal{V}(\xi)$ has contributions from the following terms:
\begin{itemize}
\item The bulk cosmological constant in \ref{Eq5}
\item The 5-d Ricci scalar: The second and third terms of \ref{Eq9} where $A^\prime$ and $A^{\prime\prime}$ are respectively obtained from \ref{6a} and \ref{2} while $V_{vis}$ and $V_{hid}$ are taken from \ref{Eq10}.
\item The brane tensions, i.e. \ref{Eq6} and \ref{Eq7}.
\item The potential and kinetic terms of the bulk scalar field given in \ref{Eq5a} where $\tilde{V}(\Psi)$ is given in \ref{7} while the kinetic term can be constructed from \ref{4} and \ref{2}.
\end{itemize}
The final form of the radion potential is given by,
\begin{align}
\mathcal{V}(\xi)=6\omega^2 h(\xi) + 6\gamma J(\xi)
\label{Eq14}
\end{align}
where $h(\xi)$ is given in \ref{Eq12} while $J(\xi)$ assumes the form,
\begin{align}
\label{Eq15}
J(\xi)=\frac{c_2^4}{16}\Bigg[\xi^4 -1 + 2 \frac{\omega_m^6}{c_2^6} \Bigg\lbrace \frac{1}{\xi^2}-1
\Bigg\rbrace - \frac{\omega_m^8}{c_2^8} \Bigg\lbrace \frac{1}{\xi^4}-1\Bigg\rbrace - 2\frac{\omega_m^2}{c_2^2} \Bigg\lbrace {\xi^2}-1\Bigg\rbrace\Bigg]
\end{align}

The above calculation shows that the presence of non-flat branes and the bulk scalar field naturally give rise to a potential for the radion in the 4-d effective action which in turn can stabilize the modulus. This is particularly interesting, as earlier attempts to stabilize the modulus invokes a bulk scalar field whose origin remains unexplored \cite{Goldberger:1999uk}. In the present work, the higher dimensional scalar owes its origin to the higher curvature degrees of freedom in the bulk. With curvature as high as the Planck scale in the bulk, higher curvature corrections to the gravity action may be expected. It may however be noted that even with bulk Einstein gravity (and no scalar field), a stabilizing potential for radion can be generated in the 4-d effective action if the branes are allowed to be non-flat. We have explored this earlier \cite{Banerjee:2017jyk,Banerjee:2018kcz,Banerjee:2020uil} and such a study reveals that in the absence of the bulk scalar the forms of the non-minimal coupling (\ref{Eq12}), the non-canonical kinetic term (\ref{Eq13}) and the first term of $\mathcal{V}(\xi)$ (\ref{Eq14}) remain the same with $\omega_m\to \omega$ while the last term in \ref{Eq14} vanishes. 
In the event $\omega\to 0$ and $\gamma\to 0$ the radion potential identically vanishes and the radion kinetic term becomes canonical. 
The role of the bulk scalar (which originates from 5-d $f(R)$ gravity) therefore primarily lies in modifying the form of the radion potential.

\subsection{Tranformation to the 4-d Einstein frame}
\label{S3-2}
From the previous section it is clear that the effective action is obtained in the Jordan frame since the Ricci scalar has a non-minimal coupling with the radion field \cite{Karam:2018squ}. The goal of this section is to express the effective action in the Einstein frame through a conformal transformation of the metric $\tilde{g}_{\mu\nu}$ \cite{PhysRevD.65.023521,RevModPhys.82.451,DeFelice2010,PhysRevD.92.026008,0264-9381-14-12-010,BARROW1988515,1475-7516-2005-04-014,PhysRevD.76.084039,1475-7516-2013-10-040}, i.e. we define a new metric $g_{\mu\nu}$ such that $g_{\mu\nu}=\Upsilon^2(x) \tilde{g}_{\mu\nu}$ where $g^{\mu\nu}=\Upsilon^{-2}(x) \tilde{g}^{\mu\nu}$ and $\sqrt{-\tilde{g}}=\Upsilon^{-4}(x)\sqrt{-g}$.
With the above transformation it can be shown that the Ricci scalar in the Einstein frame is related to that of the Jordan frame by,
\begin{equation}
R=\left[\frac{\tilde{R}}{\Upsilon^2}
-\frac{6}{\Upsilon^3}\tilde{g}^{\mu\nu}\tilde{\nabla}_\nu\tilde{\nabla}_\mu\Upsilon \right]
\label{Eq16}
\end{equation}
in four dimensions.
Further, by choosing $\Upsilon \equiv \sqrt{h(\xi)}$ the non-minimal coupling to the Ricci scalar can be removed, while the non-canonical coupling to the kinetic term and the radion potential gets modified as compared to that of the Jordan frame. 
The effective action in the Einstein frame can be written as,


\begin{align}
S^{\rm E}=\int d^4 x \sqrt{-{g}}\Bigg[\frac{2M^3}{k}R
-\frac{6M^3c_2^2}{k}\frac{1}{2}P(\xi)\partial^\mu\xi\partial_\mu\xi
-2M^3k V(\xi)\Bigg]  
\label{Eq17} 
\end{align}
where, ${P}(\xi)$ is given by,
\begin{align}
P(\xi)&=\frac{\mathcal{G}(\xi)}{h(\xi)}+\frac{1}{c_2^2}\bigg[\frac{h^\prime(\xi)}{h(\xi)}\bigg]^2  
\label{Eq18}   
\end{align}
while $V(\xi)$ assumes the form,
\begin{align}
V(\xi)=\frac{\mathcal{V}(\xi)}{h(\Phi/f)^2}=\frac{6\omega^2}{h(\xi)} + \frac{6\gamma J(\xi)}{h(\xi)^2}~,
\label{Eq19} 
\end{align}
with $h(\xi)$ and $J(\xi)$ are shown in \ref{Eq12} and \ref{Eq15} respectively.

\begin{figure}[t]
\centering
\includegraphics[scale=0.5]{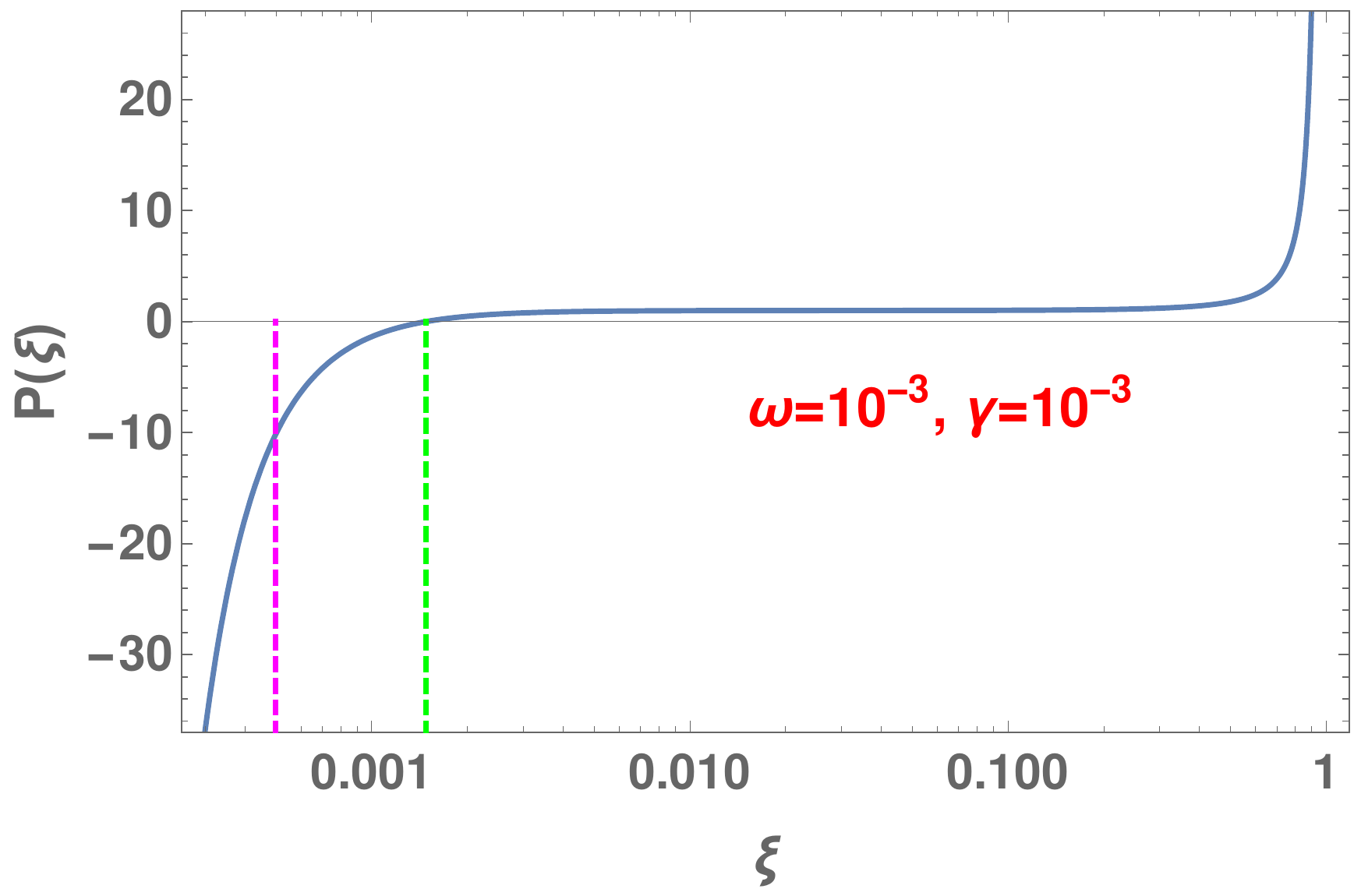}~~
\caption{The above figure depicts the variation of $P(\xi)$ with $\xi$ for $\omega=10^{-3}=\gamma$. The non-canonical coupling to the kinetic term has a zero crossing at $\xi=\xi_c$ denoted by the green dashed line. The magenta dashed line represents the value of $\xi$ where the potential has an inflection point.
For more details see text. }
\label{Fig_01}
\end{figure}

The non-canonical coupling to the kinetic term $P(\xi)$ and the radion effective potential $V(\xi)$ exhibit several interesting features which we now discuss.
In \ref{Fig_01} we plot the variation of $P(\xi)$ with $\xi$ for the representative case $\omega=10^{-3}$. In general $P(\xi)$ has a zero crossing at $\xi=\xi_c $ (shown with the green dashed line in see \ref{Fig_01}) such that when $\xi<\xi_c$ the radion field is in the phantom regime. The value of $\xi_c$ depends primarily on $\omega$ (which characterizes the non-flatness of the branes through the presence of the brane cosmological constant). For example, with $\omega=10^{-3}$,  
$\xi_c\simeq 0.0014838, ~0.0015576, {~\rm and}~ 0.0122798$,  for $\gamma=10^{-3},~0.1~ {\rm and}~0.99$ respectively. For values of $\gamma<10^{-3}$, $\xi_c\to 0.00148$ for $\omega=10^{-3}$. This behavior of $P(\xi)$ is quite generic and holds good for other values of $\omega$ as well. 
For completeness we also mention that for high values of $\omega$, (e.g. $\omega\simeq 1$) and near extremal $\gamma$, i.e. $\gamma\simeq 0.99$, $P(\xi)$ is always negative.

\begin{figure}[t!]
\centering
{\hspace{-1.5cm}\subfloat[\label{Fig_3a}]{\includegraphics[scale=0.42]{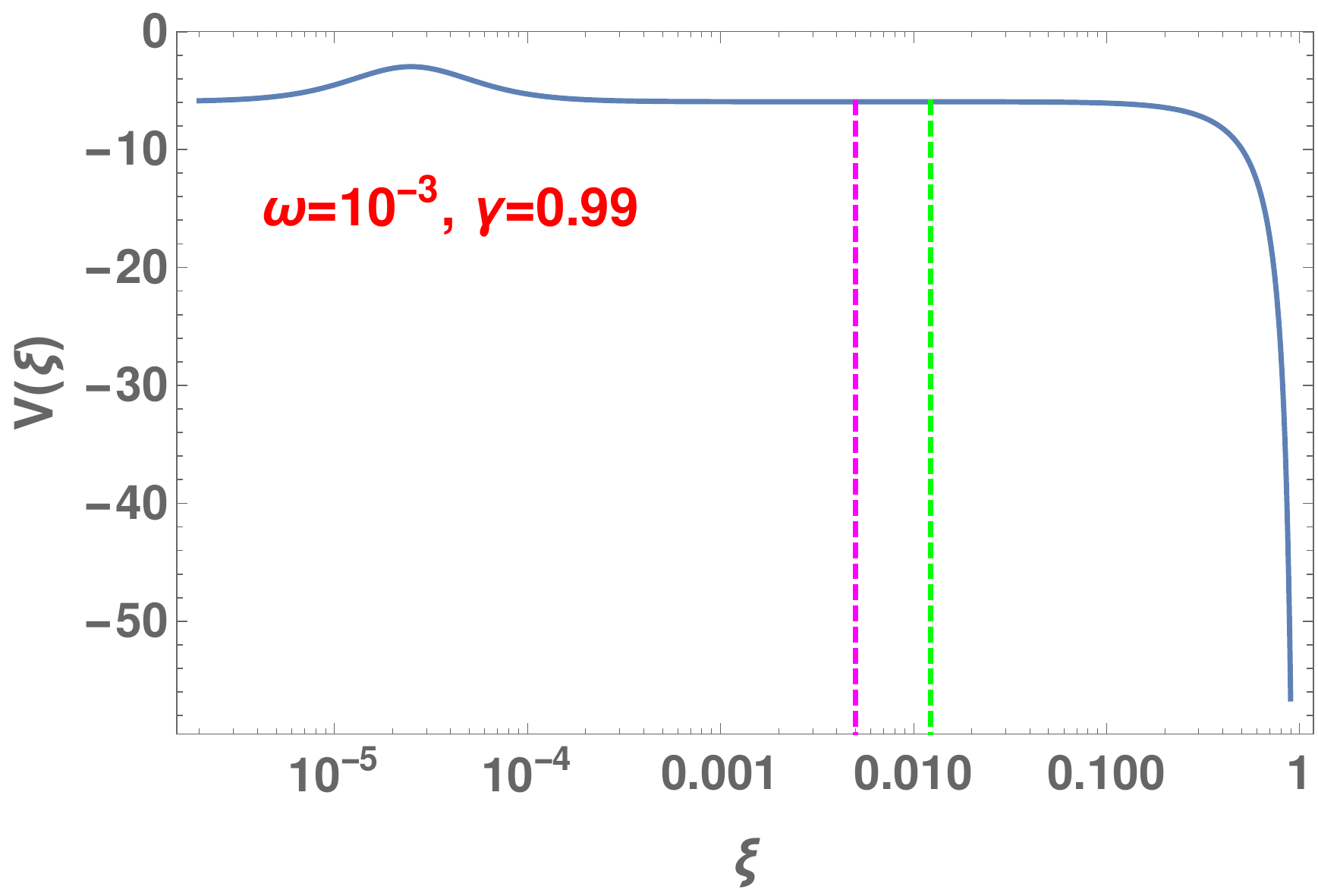}}~~
\hspace{2.cm} \subfloat[\label{Fig_3b}]{\includegraphics[scale=0.45]{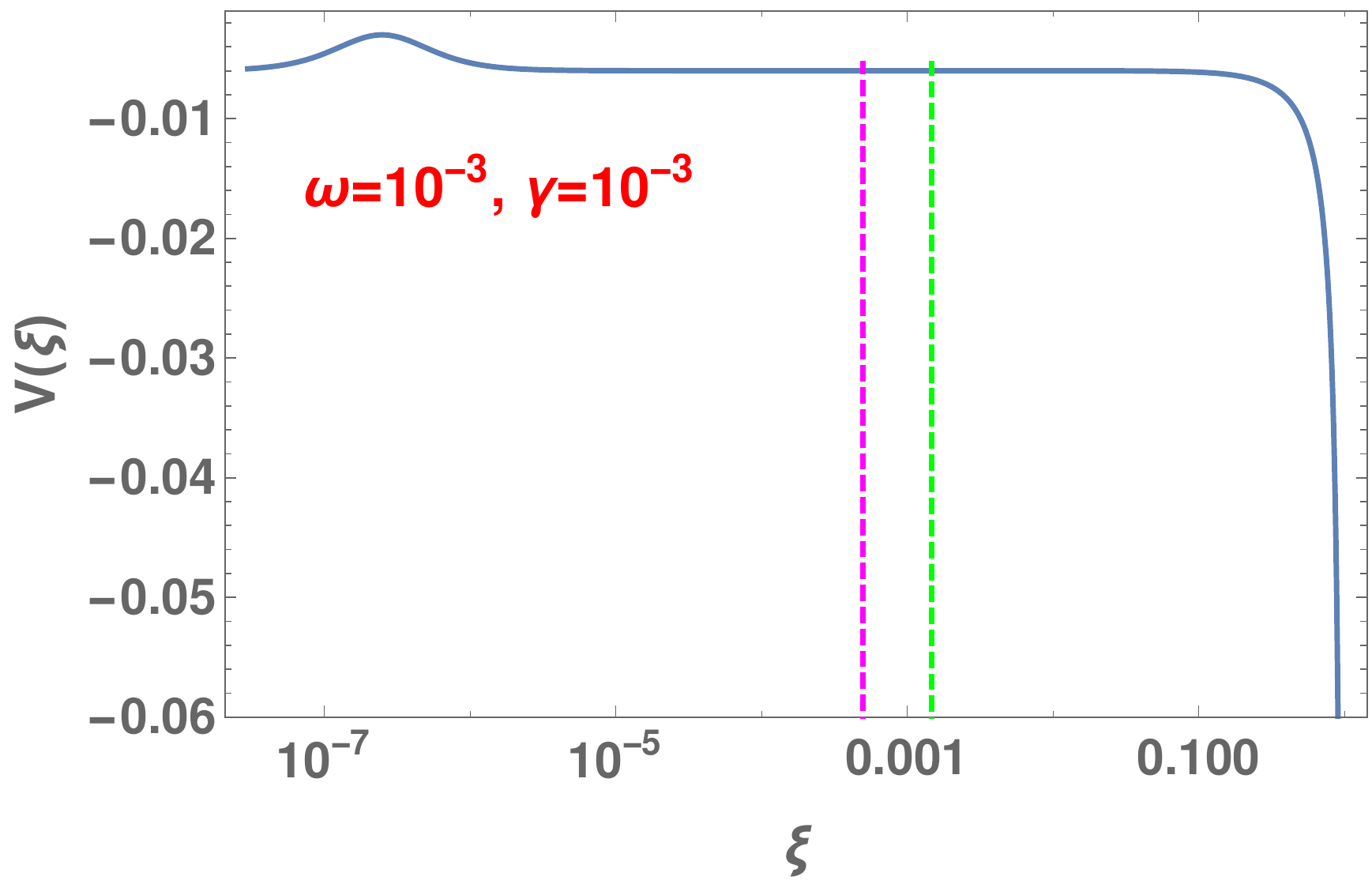}}}\\
{\hspace{-2.cm}\subfloat[\label{Fig_3c}]{\includegraphics[scale=0.48]{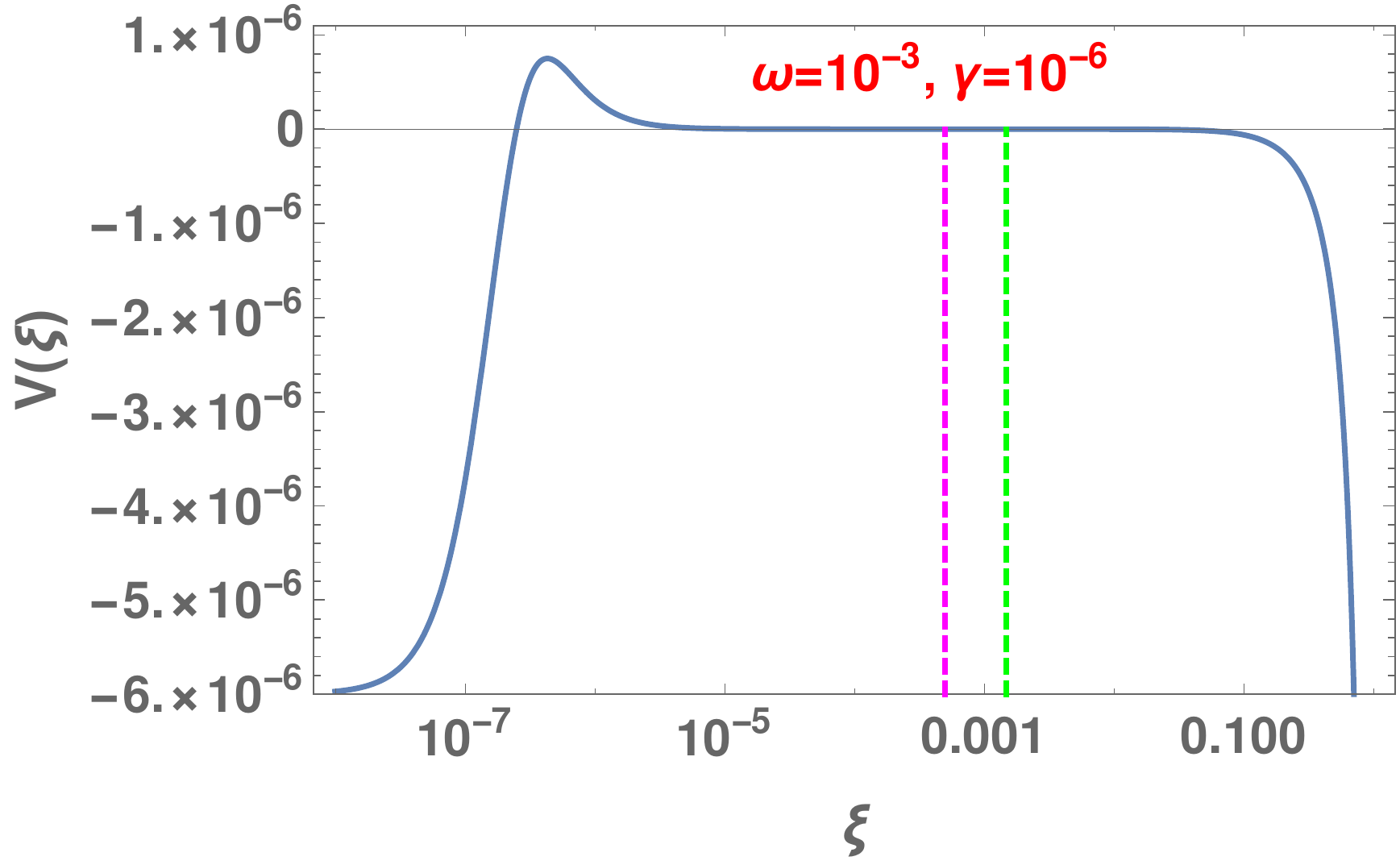}}}~~
\caption{The above figure illustrates the behavior of the radion potential $V(\xi)$ for $\omega=10^{-3}$ with (a) $\gamma=0.99$, (b) $\gamma=10^{-3}$ and (c) $\gamma=10^{-6}$. We note that $V(\xi)$ has an inflection point at $\xi=\xi_1=\frac{\omega_m}{c_2}$ (marked by the magenta dashed line). The green dashed line denotes the value of $\xi$ where the the non-canonical kinetic coupling $P(\xi)$ has a zero crossing.
The parameter $\gamma$ is directly related to the strength of the scalar field whose presence gives rise to a maxima in $V(\xi)$ which vanishes with smaller and smaller values of $\gamma$ (\ref{Fig_3a}---\ref{Fig_3c}).  
}
\label{Fig_03}
\end{figure}

Next we discuss the main characteristics of the radion effective potential $V(\xi)$. We note from \ref{Fig_03} that $V(\xi)$ has an inflection point at $\xi=\xi_1={\omega_m/c_2}$ (marked by the magenta dashed line in \ref{Fig_01} and \ref{Fig_03}) and a maxima (see \ref{Fig_03}).
We discuss analytically the characteristics of $V(\xi)$ in various regimes of $\xi$ in the Appendix.
We recall that $V(\xi)$ is given by \ref{Eq19} such that the presence of the bulk scalar has two fold effect on the radion effective potential: 
(i) modify the brane cosmological constant from $\omega$ to $\omega_m$ and (ii) introduce the second term in \ref{Eq19} which is directly 
proportional to the strength of the scalar field $\gamma$. Therefore higher values of $\gamma$ make the second term in \ref{Eq19} 
dominant compared to the first term. Note that for a given $\omega$, $\gamma>0$ while $J(\xi)<0$, which implies that for 
$\gamma\geq \frac{\omega}{c_2}$, $V(\xi)$ is always negative as shown in \ref{Fig_3a} and \ref{Fig_3b}. 
The position of the inflection point corresponds to $\xi_1\simeq \frac{\omega_m}{c_2}$ while the position of the maxima almost coincides with $\xi= \xi_2\simeq \omega_m^2/c_2^2$ (when $\gamma\lesssim \omega$). These results are analytically derived in the Appendix. The zero crossing of the kinetic term  is represented by the point where the green dashed line touches $V(\xi)$ in \ref{Fig_03}. 
When $\gamma\simeq \omega^2$ the first term in \ref{Eq19} starts dominating over the second term and $V(\xi)$ becomes positive near the maxima. With smaller and smaller values of $\gamma$ the maxima disappears, i.e. $V(\xi)$ becomes positive for all $0<\xi<1$ (i.e. in the physically allowed range) and has only the inflection point.

It may also be noted that $\xi_1,\xi_2 \leq \xi_c$, i.e., the maximum 
and the inflection points of the potential lie in the phantom regime which has interesting consequences pertaining to the stabilization of the modulus. One can further show that the hierarchy problem may be addressed (i.e. $e^{-A}\simeq 10^{-16}$) when the radion is at its stable point for appropriate choice of $\omega_m$.
An inflection point in the potential often plays an important role in the early universe scenario 
\cite{Okada:2017cvy,Okada:2016ssd,Dimopoulos:2017xox,Baumann:2007np,Baumann:2007ah,Allahverdi:2006iq,Allahverdi:2006we,BuenoSanchez:2006rze}. 
Moreover, scalar field potentials with inflection points are particularly interesting for the production of primordial 
black holes in the early universe \cite{Braglia:2020eai}.  
The presence of a phantom regime of the modulus naturally leads to a violation of the null energy condition, making this a viable 
model to study non-singular bounce \cite{Banerjee:2020uil}.
Moreover, since the kinetic term is negative when $\xi$ attains its maximum, $\xi_2$ corresponds to the stable point of the radion.

We would like to address the modulus stabilization in the conformally connected F(R) model where the action 
is given by \ref{transformation1}, with the form of F(R) is obtained in \ref{form of F 4}. 
Solutions of metric for this F(R) model can be extracted from the solutions of corresponding scalar-tensor theory 
by using the inverse conformal transformation of the metric \ref{1}. Therefore the line element in the F(R) frame turns out to be,
\begin{eqnarray}
 ds^2 = e^{-\frac{\kappa}{\sqrt{3}}\left(\Psi(\varphi) - \Psi_P\right)}\left[e^{-2A(\varphi)}\tilde{g}_{\mu\nu}(x)dx^{\mu}dx^{\nu} + r_c^2d\varphi^2\right]~,
 \label{final1 }
\end{eqnarray}
where $\Psi(\varphi)$ is the bulk scalar field and determined in \ref{10}. Here it may be mentioned that due to the complicated form of 
F(R), it seems difficult to obtain the exact form of the modulus potential in the F(R) frame, and thus we will determine the same 
in the leading order of the brane cosmological constant, by considering $\omega^2 \ll 1$. 
Such consideration is indeed viable, as the $\omega^2$ should be small in order to solve the gauge hierarchy problem. 
The bulk Ricci scalar, in the leading order of $\omega^2$, becomes $R = 4k^2\left(3\omega^2e^{A(\varphi)} - 5\right)$. Due to this expression 
of Ricci scalar, the form of F(R), from \ref{form of F 4}, takes the following form,
\begin{eqnarray}
 F(R)&=&\omega^2e^{2kr_c\varphi}\left[24k^2M^3~e^{38/\left(9\gamma\right)}\left(1 - 
 \frac{Q\left(\gamma,\Lambda,\omega\right)}{2P\left(\gamma,\Lambda,\omega\right)}\right)
 \left(1-\frac{10}{9\gamma P\left(\gamma,\Lambda,\omega\right)}\right) 
 - \frac{152\Lambda}{27\gamma^2P^2\left(\gamma,\Lambda,\omega\right)}\right]\nonumber\\ 
 &-&\left[40k^2M^3~e^{38/\left(9\gamma\right)}\left(1 - 
 \frac{Q\left(\gamma,\Lambda,\omega\right)}{2P\left(\gamma,\Lambda,\omega\right)}\right) 
 + \frac{\Lambda}{P^2\left(\gamma,\Lambda,\omega\right)}\left(\frac{38}{9\gamma}\right)^2\right]~,
 \label{final2}
\end{eqnarray}
with $\Lambda = -24k^2M^3$ is rhe bulk cosmological constant, and $P\left(\gamma,\Lambda,\omega\right)$, $Q\left(\gamma,\Lambda,\omega\right)$ 
are shown in \ref{P and Q}. Plugging back the above form of F(R) into action \ref{transformation1} and integrating the extra dimensional 
coordinate yields the effective forur dimensional modulus potential as follows,
\begin{eqnarray}
 V_\mathrm{eff}(r_c)&=&M^3k\bigg\{\left[5~e^{38/\left(9\gamma\right)}\left(1 - 
 \frac{Q\left(\gamma,\Lambda,\omega\right)}{2P\left(\gamma,\Lambda,\omega\right)}\right) 
 - \frac{6}{P^2\left(\gamma,\Lambda,\omega\right)}\left(\frac{38}{9\gamma}\right)^2\right]\bigg(1 - e^{-4kr_c\pi}\bigg)\nonumber\\
 &-&\omega^2\left[12~e^{38/\left(9\gamma\right)}\left(1 - 
 \frac{Q\left(\gamma,\Lambda,\omega\right)}{2P\left(\gamma,\Lambda,\omega\right)}\right)
 \left(1-\frac{10}{9\gamma P\left(\gamma,\Lambda,\omega\right)}\right) 
 + \frac{608}{9\gamma^2P^2\left(\gamma,\Lambda,\omega\right)}\right]\bigg(1 - e^{-2kr_c\pi}\bigg)\bigg\}~~.
 \label{final3}
\end{eqnarray}
Therefore the modulus potential can be expressed as,
\begin{eqnarray}
 V_\mathrm{eff}(r_c) = M^3k\bigg\{\beta\left(1 - e^{-4kr_c\pi}\right) - \omega^2\alpha\left(1-e^{-2kr_c\pi}\right)\bigg\}~,
 \label{final4}
\end{eqnarray}
where $\beta$ and $\alpha$ symbolize the coefficients of the respective terms. Due to $0 < \gamma < 1$, we get $P \approx -\frac{13}{3\gamma}$ and 
$Q \approx -2$ respectively (from \ref{P and Q}), in effect of which, both the coefficients $\alpha$ and $\beta$ turn out to be positive. 
As a result, $V_\mathrm{eff}(r_c)$ gets a maximum at $\langle r_c \rangle$, where,
\begin{eqnarray}
 \exp{\left(-\pi\langle  kr_c \rangle\right)} = \omega\sqrt{\frac{\alpha}{\beta}}~.
 \label{final5}
\end{eqnarray}
Therefore similar to the scalar-tensor theory, the modulus potential in the F(R) frame gets a maximum for $\omega^2 \ll 1$. However due to the 
conformal factor connecting the two theories, the stable value of $k\pi r_c$ in the F(R) model becomes different 
compared to that in the scalar-tensor frame. Moreover, \ref{final4} clearly indicates that for $\omega = 0$ (i.e when the branes are flat), the 
modulus potential is not stablized and the branes get separated to infinity. Thus if the branes are flat (i.e. $\omega=0$) then the higher curvature gravity in this model fails to stabilize the modulus. However, in presence of non-flat branes in F(R) model we obtain a global stable point for the modulus in contrast to an inflection point in an Einstein bulk with non-flat branes.

Before concluding, we would like to mention that in case the bulk action is governed by Einstein or $f(R)$ gravity and 
the 3-branes are non-flat a potential for the radion field arises naturally at the level of 4-d effective action which has an inflection point. 
The underlying reason for the presence of the inflection point is the non-flat nature of the 3-branes. 
The inflection point occurs at $\xi=\xi_1=\omega_m/c_2$ where $\omega_m \propto \Omega$, the brane cosmological constant, 
when the radion is in the phantom regime. Such a result has several interesting consequences: (a) It leads to a violation of the null energy 
condition that naturally leads to non-singular bounce (b) Scalar field potentials with inflection
points are particularly interesting for the production of primordial black holes in the early universe (c) In case of $f(R)$ gravity, 
apart from the inflection point the radion also has a maximum in the phantom regime which in turn corresponds to the stable point of the radion.

\section{Conclusion}
\label{Conclusion}

In this work we explore the possibility of modulus stabilization in a bulk F(R) gravity model with non-flat branes sitting at the orbifold fixed points. Such stabilization is known to be crucial to resolve the gauge hierarchy problem via warped geometry models. The role of an external bulk scalar in achieving this stabilization has been explored by several authors in the presence  of Minkowskian 3-branes in an Einsteinian bulk.
With bulk curvature as high as the Planck scale a non-trivial contribution of the higher curvature corrections to Einstein grvity is inevitable. It is well known that an F(R) gravity model is equivalent to an Einstein gravity with a scalar field where the scalar field encapsulates the additional degree of freedom of the higher curvature.  Efforts to look for modulus stabilization through such a \emph{naturally} originating bulk scalar became an important area of study. Though  some of the earlier works attempted to address this issue, none of these could come up with an exact analytic solution for the bulk geometry by incorporating the back-reaction of the scalar in presence of  non-flat branes at the boundaries. This has been addressed in this work for the first time. 

By introducing non-flat branes in a bulk F(R) gravity we study the nature of the radion field in the effective 4-d action.
Our findings bring out the following remarkable features in the context  of radion stabilization process : 
\begin{itemize}
\item The higher curvature terms in the bulk as well as non-flat nature of branes play crucial role in stabilizing  the modulus without the need of any external field. 
\item The form of F(R) that leads to an exact solution of the bulk geometry bears all positive integer powers of R and hence is physically well motivated.
\item Non-flat branes along with the higher curvaure terms in the bulk generate modulus potential and a  non-canonical kinetic term for radion in the 4-d effective action.
\item In the process of approaching it's stable value , the radion field  passes through  a phantom regime. This points towards a possibility that due to the violation of the  null energy condition  such a radion field  may cause  a non-singular bounce for the Universe without the need of invoking any other field to generate the bounce.
\item As the radion approaches its stable value the resulting modulus consistently resolves the gauge hierarchy problem in 4-d effective theory for a corresponding  value of the brane cosmological constant 
$\omega$ determined by the model.
\end{itemize}
This work thus brings out the remarkable role of the higher curvature geometry and brane curvature to determine the character and stabilization of the  modulus field (i.e. the radion) in a non-flat warped geometry model where the geometry itself stabilizes the modulus. Moreover, in addition to addressing the gauge-hierarchy problem it brings out the possibility of a non-singular bounce to address the long-standing problem of cosmological singularity through the resulting modulus potential. Furthermore, the presence of the inflection point in the radion potential makes this a viable model to study the formation of primordial black holes. We propose to address these issues in a future work.

\section*{Appendix}
In order to determine the extrema of $V(\xi)$ (if any), we re-define 
the radion field as,
\begin{eqnarray}
 \xi = \xi_\mathrm{0}\left(1 + \epsilon\right)~~~~~~~~~~~~~,~~\mathrm{with}~~~~~~~~~~\epsilon \ll 1~~,
 \label{extremum 1}
\end{eqnarray}
where $\xi_\mathrm{0}$ being a constant, and thus $\epsilon$ is equivalently treated as radion field. With such redefinition of the radion field, 
we may divide the entire range of $\xi$ by three sections: (1) $\xi_\mathrm{0} = \omega_m/c_2$, i.e when the radion field lies around the value 
$\omega_m/c_2$, (2) $\xi_\mathrm{0} \ll \omega_m/c_2$, or equivalently, when $\xi$ lies far below than $\frac{\omega_m}{c_2}$, and (3) 
$\xi_\mathrm{0} \gg \omega_m/c_2$, which covers when $\xi$ is much larger than $\frac{\omega_m}{c_2}$. Below, we 
consider these three cases separatley to cover the entire range of the radion field. However, before doing so, the quantities $h(\xi)$ and 
$J(\xi)$, in terms of $\epsilon$, are determined as,
\begin{eqnarray}
 h(\epsilon) = \omega_m^2\left[\frac{c_2^2}{4\omega_m^2}\big(1 - \xi_\mathrm{0}^2\left(1 + \epsilon\right)^2\big) 
 - \frac{\omega_m^2}{4c_2^2}\left(1 - \frac{1}{\xi_\mathrm{0}^2\left(1 + \epsilon\right)^2}\right) 
 + \ln{\left[\xi_\mathrm{0}\left(1 + \epsilon\right)\right]}\right]
 \label{extremum 2}
\end{eqnarray}
and
\begin{eqnarray}
 J(\epsilon) = &-&\frac{c_2^4}{16}\bigg[1-\xi_\mathrm{0}^4\left(1 + \epsilon\right)^4 - 2\left(\frac{\omega_m}{c_2}\right)^2
 \left(1-\xi_\mathrm{0}^2\left(1 + \epsilon\right)^2\right) + 2\left(\frac{\omega_m}{c_2}\right)^6\left(1 - 
 \frac{1}{\xi_\mathrm{0}^2\left(1 + \epsilon\right)^2}\right)\nonumber\\ 
 &-&\left(\frac{\omega_m}{c_2}\right)^8\left(1 - 
 \frac{1}{\xi_\mathrm{0}^4\left(1 + \epsilon\right)^4}\right)\bigg]
 \label{extremum 3}
\end{eqnarray}
respectively.

\subsection*{Case-I: $\xi_\mathrm{0} = \frac{\omega_m}{c_2}$} 
 In this case, plugging the above expressions of $h(\epsilon)$ and 
 $J(\epsilon)$ into \ref{Eq19}, and expanding the radion potential (i.e $V(\epsilon)$) in a Taylor series around $\epsilon = 0$, we get,
 \begin{eqnarray}
  V(\epsilon) = \frac{6\left\{\left(1-\frac{\omega_m^4}{c_2^4}\right)\left(-\gamma + \frac{4\omega_m^2}{c_2^2} + 6\gamma~\frac{\omega_m^2}{c_2^2}\right) 
  + 16\left(\frac{\omega_m}{c_2}\right)^4(1-\gamma)\ln{\left(\frac{\omega_m}{c_2}\right)}\right\}}
  {\left\{1 + 4\left(\frac{\omega_m}{c_2}\right)^2\ln{\left(\frac{\omega_m}{c_2}\right)} - \left(\frac{\omega_m}{c_2}\right)^4\right\}^2} 
  + \mathcal{O}\left(\epsilon^3\right)~,
  \label{extremum 4}
 \end{eqnarray}
where it may be noticed that the first and second order terms of the Taylor series identically vanish at $\epsilon=0$, in particular,
\begin{eqnarray}
 \frac{dV}{d\epsilon}\bigg|_{\epsilon = 0} = 0~~~~~~~\mathrm{and}~~~~~~~\frac{d^2V}{d\epsilon^2}\bigg|_{\epsilon = 0} = 0~.\nonumber
\end{eqnarray}
 However the term of $\mathcal{O}\left(\epsilon^3\right)$ contributes in the expression of $V(\epsilon)$. Therefore we may argue that the radion potential 
 has an $inflection~point$ at $\xi = \frac{\omega_m}{c_2}$. Moreover for a small value of brane cosmological constant, in particular for 
 $\frac{\omega_m}{c_2} \ll 1$, \ref{extremum 4} immediately leads to the zeroth order term of $V(\epsilon)$ 
 (i.e the term of $\mathcal{O}\left(\epsilon^0\right)$) as $\approx -6\gamma$, which is indeed negative (as $\gamma > 0$). As a result, 
 the radion potential gets negative at the point of inflection. Here we would like to mention that the consideration 
 $\frac{\omega_m}{c_2} \ll 1$ has phenomenological motivation, mainly, regarding the resolution of the gauge hierarchy problem.

 \subsection*{Case-II: $\xi_\mathrm{0} \ll \frac{\omega_m}{c_2}$}
 In this case, with the help of $h(\epsilon)$ and $J(\epsilon)$ 
 (see \ref{extremum 2} and \ref{extremum 3}), we expand the radion potential in a Taylor series around $\epsilon = 0$, as follows, 
 \begin{eqnarray}
  V(\epsilon) = V_\mathrm{0} + \epsilon V_\mathrm{1} + \epsilon^2 V_\mathrm{2} + \mathcal{O}\left(\epsilon^3\right)~,
  \label{extremum 5}
 \end{eqnarray}
where the zeroth order term, i.e $V_\mathrm{0}$ has the following expression,
 \begin{eqnarray}
  V_\mathrm{0} = \frac{6\omega^2}{h(\epsilon)} + \frac{6\gamma J(\epsilon)}{h(\epsilon)^2}\bigg|_{\epsilon = 0}~.
  \label{extremum 6}
 \end{eqnarray}
 Moreover the first order term, i.e $V_\mathrm{1}$ is given by,
 \begin{eqnarray}
  V_\mathrm{1} = 2\left(\frac{\omega_m}{c_2}\right)^6 - 2\xi_\mathrm{0}^2\left(\frac{\omega_m}{c_2}\right)^2 
  + \gamma\left\{\xi_\mathrm{0}^2 - \left(\frac{\omega_m}{c_2}\right)^4 - 4\left(\frac{\omega_m}{c_2}\right)^6\ln{\xi_\mathrm{0}} 
  + \left(\frac{\omega_m}{c_2}\right)^8\right\}~,
  \label{extremum 7}
 \end{eqnarray}
 which is found to have a zero crossing at,
 \begin{eqnarray}
  \xi_\mathrm{0} \approx \left(\frac{\omega_m}{c_2}\right)^2\left[1 - 4\left(\frac{\omega_m}{c_2}\right)^2\ln{\left(\frac{\omega_m}{c_2}\right)}\right]~.
  \label{extremum 8}
 \end{eqnarray}
 Consequently, at such zero crossing point, the second order term becomes negative, in particular,
 \begin{eqnarray}
  V_\mathrm{2} = 6\left(\frac{c_2}{\omega_m}\right)^2\left[-\gamma + \mathcal{O}\left(\frac{\omega_m^2}{c_2^2}\right)\right] 
  \approx -6\gamma\left(\frac{c_2}{\omega_m}\right)^2 < 0~~.
  \label{extremum 9}
 \end{eqnarray}
 Thereby at $\xi \approx \left(\frac{\omega_m}{c_2}\right)^2\left[1 - 4\left(\frac{\omega_m}{c_2}\right)^2\ln{\left(\frac{\omega_m}{c_2}\right)}\right]$, 
 which indeed satisfies $\xi \ll \frac{\omega_m}{c_2}$, the first order term vanishes and consequently, the radion potential can be written as,
 \begin{eqnarray}
  V(\epsilon) = V_\mathrm{0} + \epsilon^2 V_\mathrm{2} + \mathcal{O}\left(\epsilon^3\right)\nonumber
 \end{eqnarray}
with $V_\mathrm{2} < 0$. This indicates that the radion potential gets a maximum at a value of $\xi$ as mentioned in \ref{extremum 8}.
 
\subsection*{Case-III: $\xi_\mathrm{0} \gg \frac{\omega_m}{c_2}$}
In this regime of $\xi$, the corresponding radion potential can be 
expanded around $\epsilon = 0$ as,
\begin{eqnarray}
 V(\epsilon) = V_\mathrm{0} + \epsilon W_\mathrm{1} + \mathcal{O}\left(\epsilon^2\right)~,
 \label{extremum 10}
\end{eqnarray}
where $V_\mathrm{0}$ has the same expression as \ref{extremum 6}, however with $\xi_\mathrm{0} \gg \frac{\omega_m}{c_2}$. The first order term, i.e 
$W_\mathrm{1}$, is given by,
\begin{eqnarray}
 W_\mathrm{1} = -\gamma\left[1 + 4\left(\frac{\omega_m}{c_2}\right)^2\right]~,
 \label{extremum 11}
\end{eqnarray}
which, clearly, has no zero crossing for any possible value of $\xi$ (as $\gamma \neq 0$). This argues that, in the regime 
$\xi \gg \frac{\omega_m}{c_2}$, the radion potential has $no~extremum$ at all. Moreover \ref{extremum 11} depicts that 
$W_\mathrm{1}$ is negative, which clearly demonstrates that the radion potential decreases with increasing value of the radion field.

\bibliography{radioninflation}
\bibliographystyle{./utphys1}

\end{document}